\documentclass[twocolumn,prb,showpacs,floatfix]{revtex4}

\usepackage{amssymb}
\usepackage{amsmath}
\usepackage{epsfig}
\usepackage{graphics}
\usepackage[dvips]{color}

\begin{document}

\title{Graphene in a Strong Magnetic Field: Massless Dirac Particles
vs. Skyrmions}

\author{John Schliemann}

\affiliation{Institute for Theoretical Physics, University of 
Regensburg, D-93040 Regensburg, Germany}

\date{\today}

\begin{abstract}
We discuss models for massless Dirac fermions being subject to a 
perpendicular magnetic field in spherical geometry. These models are
analogues of Haldane\rq s spherical construction for massful charge
carriers. The single particle states constructed here are easily
implemented in existing numerical code for many-body problems in
conventional quantum Hall systems. Moreover, the many-body states of
fully filled sublevels in the subspace of lowest Landau level index
are skyrmions with respect to the layer spin. 
\end{abstract}

\maketitle

\section{Introduction}

In the recent years, graphene has developed to one of clearly most
active directions of work in today\rq s both experimental
and theoretical condensed matter
physics \cite{Novoselov04,Geim07,CastroNeto07}. Compared to conventional
two-dimensional electronic systems, the peculiar properties 
of graphene mainly stem from its linear dispersion near the Fermi energy,
and the chiral nature of electronic states entangling the momentum and
sublattice degree of freedom \cite{Geim07,CastroNeto07}.
Among a plethora of interesting phenomena, the
quantum Hall effect occurring at anomalous (``half-integer'') filling factors
is one of the most spectacular observations in this new type of
material \cite{Novoselov05,Zhang05,Novoselov07}.
Other partially related unusual features of graphene include a 
cyclotron mass being
proportional to the square root of the particle density 
\cite{Novoselov05}, and a non-equidistant Landau level spectrum 
proportional to the square root
of the magnetic field
\cite{Novoselov05,Zhang05,Novoselov07,Deacon07,Jiang07a,Jiang07b}. 

In quantum Hall physics, the exact numerical treatment of finite many-body
systems has always been an important source of theoretical evidence
\cite{Yoshioka02}. A particularly convenient model for such numerical
simulations was given already 25 years ago by Haldane \cite{Haldane83}.
In this construction, massful electrons move on a spherical surface
in a radial monopole magnetic field. This model has the advantage of lacking
any system edge (and therefore reducing finite-size effects in numerical 
results). On the other hand, the mathematical properties of the states in the 
lowest Landau level are particularly simple, enabling also substantial 
analytical progress. Both aspects have made the Haldane sphere to a widely used
model in theoretical, in particular numerical descriptions of 
quantum Hall systems;
for some representative references see \cite{Haldane85,Fano86,Haldane88a,Haldane88b,Rezayi91,Moon95,Wu95,Xie96,Yang96,Morf98,Schliemann00,Schliemann01,Wojs02,Morf02,Schliemann03,Simon03,Feiguin08,Moller08}.
Moreover, most recently this spherical model, explicitly constructed
for massful electrons in conventional two-dimensional systems, was
also applied to the situation of {\em massless} carriers in
graphene \cite{Apalkov06,Toke06,Toke07a,Toke07b,Shibata08}. In particular, 
the authors of Ref.~\cite{Toke07a} state
that ``the analogous solution for carriers with linear dispersion ... is not
known for the spherical geometry''.  The purpose of this note is 
introduce such models of massless charge carriers in the spherical geometry.
As we shall see below, they are rather straightforwardly developed
from the situation of massful particles. A particular feature of these
models are many-body ground states having the features of skyrmions
known from conventional quantum Hall monolayers.
Finally we mention previous work by other authors on Dirac fermions coupled to 
gauge fields in spehrical geometry inspired by
fullerrenes\cite{Osipov00,Kolesnikov06,Pudlak06}, i.e. 
another allotrope of carbon . Differently from the
models to be discussed here, these constructions do not rely on
angular momentum operators and have therefore also different spectra.

This paper is organized as follows. In section \ref{haldane} we
review the basic properties of the spherical model for massive charge carriers
in a perpendicular magnetic field. In section \ref{massless} we briefly
recall elementary facts concerning planar graphene in a perpendicular 
magnetic field, before constructing analogous models in spherical 
geometry. Here we also discuss two-body and many-body states. We close
with conclusions in section \ref{conclusions}.

\section{Spherical model for massful charge carriers}
\label{haldane}

In Haldane\rq s spherical model for massful charge carriers in a perpendicular
magnetic field, electrons move on a spherical surface of radius $R$
which is penetrated by a radial monopole magnetic field $B=\hbar cS/eR^{2}$.
Here $e>0$ is the elementary charge, and $2S$ is the integer number of  
flux quanta $hc/e$ through the surface \cite{Haldane83}, i.e. the radius 
is given by $R=\ell\sqrt{S}$ where $\ell=\sqrt{\hbar c/eB}$ 
is the magnetic length.
The single-particle Hamiltonian reads
\begin{equation}
{\cal H}_{0}=\frac{\vec \Lambda^{2}}{2MR^{2}}=
\frac{1}{2}\omega_{c}\frac{\vec \Lambda^{2}}{\hbar S}\,.
\label{halham}
\end{equation}
Here $M$ is an effective mass, $\omega=eB/Mc$ is the cyclotron frequency,
and the kinetic angular momentum reads, using again standard notation,
\begin{equation}
\vec\Lambda=\vec r\times\left(\vec p+\frac{e}{c}\vec A\right)\,,
\end{equation}
where the vector potential yields $\nabla\times\vec A=B\hat\Omega$,
$\hat\Omega=\vec r/R$.
The operator $\Lambda$ has the elementary properties
\begin{equation}
\vec\Lambda\cdot\vec\Omega=\vec\Omega\cdot\vec\Lambda=0\,,
\end{equation}
\begin{equation}
\left[\Lambda^{\alpha},\Lambda^{\beta}\right]=i\hbar\varepsilon^{\alpha\beta\gamma}
\left(\Lambda^{\gamma}-\hbar S\Omega^{\gamma}\right)\,.
\end{equation}
In particular, the components of $\Lambda$ fail to fulfill proper
angular momentum commutation relations and can therefore not be
considered as the generators of rotations. Instead, rotations are
generated by the angular momentum operator
\begin{equation}
\vec L=\vec\Lambda+\hbar S\vec\Omega
\label{defL}
\end{equation}
fulfilling
\begin{equation}
\left[L^{\alpha},L^{\beta}\right]=i\hbar\varepsilon^{\alpha\beta\gamma}L^{\gamma}\,,
\end{equation}
\begin{equation}
\vec L\cdot\vec\Omega=\vec\Omega\cdot\vec L=\hbar S\,,
\end{equation}
\begin{equation}
\vec L^{2}=\vec\Lambda^{2}+\left(\hbar S\right)^{2}\,.
\label{squares}
\end{equation}
Thus, the eigenvalues of $\vec L^{2}$ are given by 
$\hbar^{2}l(l+1)$ with $l=S+n$ and $n\in\{0,1,2,...\}$ where $n=0$ corresponds
to the lowest Landau level. Note, however, that the spectrum
of the Hamiltonian (\ref{halham}) is not equidistant, but grows quadratically
with $n$, differently from the planar situation.  

Let us choose the usual gauge $\vec A=(\hbar S/eR)\hat\varphi\cot\vartheta$,
where $\vartheta$, $\varphi$ are the usual polar coordinates and
$\hat\varphi$ is the unit vector in the azimuthal direction
\cite{Haldane83,Fano86}. Then eigenstates in the lowest Landau level have a 
particularly simple structure given by
\begin{equation}
\varphi_{m}\left(u,v\right)
=\sqrt{\frac{2S+1}{4\pi\ell^{2}S}
\left(
\begin{array}{c}
2S \\ S+m
\end{array}
\right)}u^{S+m}v^{S-m}\,,
\label{halstate}
\end{equation}
where $u=\cos(\vartheta/2)e^{i\varphi/2}$, $v=\sin(\vartheta/2)e^{-i\varphi/2}$, and
$m\in\{-S,\dots S\}$ is the eigenvalue of $L^{z}/ \hbar$. 

We note that the gauge invariant angular momentum operator
(\ref{defL}) crucially depends on the magnetic flux and is not identically
to the gauge-dependent canonical angular momentum operator
$\vec L_{can}=\vec r\times\vec p$. Moreover, due to the relation 
(\ref{squares}), one could alternatively define the Hamiltonian
\begin{equation}
{\cal H}^{\prime}_{0}=\frac{1}{2}\omega_{c}\frac{\vec L^{2}}{\hbar S}
\label{halham2}
\end{equation}
which differs from (\ref{halham}) just by a trivial constant.

\section{Massless Dirac particles in spherical geometry}
\label{massless}

For a {\em planar} graphene sheet in a perpendicular magnetic field, the 
single-particle 
states around one of one of the two inequivalent corners of the first
Brillouin zone are described by \cite{CastroNeto07}
\begin{equation}
{\cal H}^{pl}_{(\pm)}=v\left((\pm)\pi_{x}\sigma^{x}+\pi_{y}\sigma^{y}\right)
\label{hampl1}
\end{equation}
with $\vec\pi=\vec p+e\vec A/c$ and 
$v\approx 10^{6}{\rm m/s}$. The Pauli matrices describe the sublattice 
or pseudospin degree
of freedom, and the Zeeman coupling to the physical electron spin has been 
neglected. The double sign $(\pm)$ (valley index)
determines which corner of the Brillouin zone in considered \cite{note1}. 
These two cases are, in the absence of a magnetic field,
related by time reversal  \cite{CastroNeto07} implemented by a complex
conjugation, i.e. $\sigma^{y}$ changes sign while $\sigma^{x}$ remains
unaltered. Note that this behavior is different from angular
momentum operator describing a proper spin and not a sublattice
degree o freedom. In what follows we shall concentrate on the case $(+)$.
Defining the usual bosonic operators
\begin{equation}
a=\frac{1}{\sqrt{2}}\frac{\ell}{\hbar}\left(\pi_{x}+i\pi_{y}\right)
\quad,\quad a^{+}=(a)^{+}
\end{equation}
fulfilling $[a,a^{+}]=1$, the Hamiltonian reads
\begin{equation}
{\cal H}^{pl}_{(+)}=\frac{\hbar v}{\ell}\sqrt{2}
\left(a\sigma^{-}+a^{+}\sigma^{+}\right)
\label{hampl2}
\end{equation}
where $\sigma^{\pm}=(\sigma^{x}\pm i\sigma^{y})/2$.
The well-known eigenstates
\cite{CastroNeto07} of the Hamiltonian (\ref{hampl2}) are given by 
$|0,\uparrow\rangle$ with energy $\varepsilon_{0}=0$ and, for $n>0$,
\begin{equation}
|n,\pm\rangle=\frac{1}{\sqrt{2}}\left(
|n,\uparrow\rangle\pm|n-1,\downarrow\rangle\right)
\label{eigen}
\end{equation}
with energy $\varepsilon_{n}^{\pm}=\pm(\hbar v/\ell)\sqrt{2n}$.
Here $n$ is again the Landau level index, and the arrows are obvious standard 
notation for the sublattice spin states. In particular, Landau level index  and
(sublattice) spin are entangled in these eigenstates, a feature that will be 
reproduced by the spherical models for graphene in a magnetic field to
be discussed now.

\subsection{Spherical models}

Let us now discuss models for massless Dirac particles on a sphere
penetrated by a radial monopole magnetic field, for which we shall use the same
gauge as stated before. Note that the crucial step in constructing the
spherical model for massful carriers is to replace the linear momentum of
the usual planar Landau problem with an appropriately defined angular 
momentum. This shall also be our guideline regarding the case of massless
Dirac fermions.

\subsubsection{Single-particle Hamiltonians}

Inspired by the expressions (\ref{halham}) and (\ref{halham2}) 
we consider the Hamiltonians
\begin{equation}
{\cal H}_{(\pm)}=\frac{v}{R}\vec \Lambda\cdot\vec\tau_{(\pm)}
=\frac{v}{\ell\sqrt{S}}\vec \Lambda\cdot\vec\tau_{(\pm)}
\label{hamsp}
\end{equation}
and
\begin{equation}
{\cal H}^{\prime}_{(\pm)}=\frac{v}{\ell\sqrt{S}}\vec L\cdot\vec\tau_{(\pm)}\,.
\label{hamsp2}
\end{equation}
Here the operators $\vec\tau_{(+)}$ describing the sublattice
degree of freedom are again just given by the Pauli matrices,
$\vec\tau_{(+)}=\vec\sigma$, 
while $\vec\tau_{(-)}$ are the negatives of their complex conjugates, 
$\vec\tau_{(-)}=-(\vec\sigma)^{\ast}$, in close analogy to the planar case
(\ref{hampl1}). Note that both sets of operators, although not
referring to a proper spin, fulfill the usual relations
\begin{equation}
\left[\tau^{\alpha}_{(\pm)},\tau^{\beta}_{(\pm)}\right]
=2i\varepsilon^{\alpha\beta\gamma}\tau^{\gamma}_{(\pm)}\,.
\end{equation}
This fact will chiefly facilitate the analysis of the Hamiltonian
(\ref{hamsp2}) in terms of elementary angular momentum theory. The treatment
of the Hamiltonian (\ref{hamsp}), however, is more complicated
since the components of $\vec\Lambda$ do not fulfill an angular momentum
algebra. Moreover, in the following we shall again concentrate on the case 
$(+)$, the case $(-)$ can be treated completely analogously and is just 
related via complex conjugation. 

The two Hamiltonians given in Eqs.~(\ref{hamsp2}), (\ref{hamsp}) 
differ by the radial contribution
\begin{equation}
{\cal H}^{\prime}_{(+)}-{\cal H}_{(+)}=\frac{\hbar v}{\ell}\sqrt{S}
\vec\Omega\cdot\vec\sigma\,,
\end{equation}
which is, differently from the expressions (\ref{halham2}) and (\ref{halham}),
not just a trivial constant. 
In fact, one might argue that ${\cal H}_{(+)}$ should be considered to be
closer to the planar model of graphene since the operator $\vec\Lambda$,
in contrast with $\vec L$, does not have a radial component. However, as we
shall see shortly, when concentrating on the subspace of lowest Landau level 
index $n=0$, the solutions of negative energy are simultaneous eigenstates
of $\vec L\cdot\vec\sigma$ and $\vec\Omega\cdot\vec\sigma$ (and therefore
also $\vec\Lambda\cdot\vec\sigma$) at any systems size, whereas the solutions
of positive energy become such simultaneous eigenstates in the thermodynamic
limit $S\to\infty$. Thus, as far as the subspace of lowest Landau level index 
is concerned, the eigenstates of ${\cal H}_{(+)}$ and 
${\cal H}^{\prime}_{(+)}$ are either
identical at any system size or become identical in the thermodynamic
limit, where the planar model of graphene is recovered.
Let us therefore first concentrate on the latter Hamiltonian.

Introducing a total angular momentum of the usual form
$\vec J=\vec L+\hbar\vec\sigma/2$ one can rewrite the Hamiltonian 
(\ref{hamsp2}) as
\begin{equation}
{\cal H}^{\prime}_{(+)}=\frac{v}{\ell\hbar\sqrt{S}}
\left(\vec J^{2}-\vec L^{2}-\left(\frac{\hbar}{2}\vec\sigma\right)^{2}\right)\,.
\end{equation}
Moreover, $\vec J$ commutes with the Hamiltonian and admits
total angular momentum quantum numbers $j=l\pm 1/2=S+n\pm 1/2$. Thus, the
spectrum of ${\cal H}^{\prime}_{(+)}$ reads
\begin{equation}
\varepsilon^{\prime}_{\pm}=\pm\frac{\hbar v}{\ell\sqrt{S}}
\left(S+n+\frac{1}{2}\mp\frac{1}{2}\right)\,.
\label{ergsp}
\end{equation}
Let us again focus on the lowest Landau level index $n=0$.
Using the well-known Clebsch-Gordon coefficients \cite{Sakurai94}
for coupling an
angular momentum of length $S$ with a spin $1/2$, one can explicitly
formulate the $2S+2$ eigenstates in the multiplet $j=l+1/2$, 
\begin{eqnarray}
\psi_{m}^{+}(u,v) & = & \sqrt{\frac{S+1/2+m}{2S+1}}\varphi_{m-1/2}(u,v)
|\uparrow\rangle\nonumber\\
& + & \sqrt{\frac{S+1/2-m}{2S+1}}\varphi_{m+1/2}(u,v)
|\downarrow\rangle\,,
\label{psi+}
\end{eqnarray}
where $m\in\{-S-1/2,\dots, S+1/2\}$ is the eigenvalue of $J^{z}/ \hbar$,
and the $2S$ states with $j=l-1/2$ read
\begin{eqnarray}
\psi_{m}^{-}(u,v) & = & -\sqrt{\frac{S+1/2-m}{2S+1}}\varphi_{m-1/2}(u,v)
|\uparrow\rangle\nonumber\\
& + & \sqrt{\frac{S+1/2+m}{2S+1}}\varphi_{m+1/2}(u,v)
|\downarrow\rangle
\label{psi-}
\end{eqnarray}
with $m\in\{-S+1/2,\dots, S-1/2\}$. Similarly to the planar model
of graphene, the sublattice spin and the conventional orbital motion are
entangled with each other.

Denoting $\psi_{m}^{\pm}(u,v)=\langle\vec r|m,\pm\rangle$, 
the expectation values
of $\vec\Omega\cdot\vec\sigma$ are straightforwardly calculated as
\begin{eqnarray}
\langle m,+|\vec\Omega\cdot\vec\sigma|m,+\rangle & = & \frac{2S}{2S+2}\,,
\label{spinexp+}\\
\langle m,-|\vec\Omega\cdot\vec\sigma|m,-\rangle & = & -1
\label{spinexp-}\,. 
\end{eqnarray}
Thus, the variance of $\vec\Omega\cdot\vec\sigma$ within the states
of lower energy $|m,-\rangle$ is exactly zero, while for the states 
$|m,+\rangle$ one finds
\begin{eqnarray}
 & & \langle m,+|\left(\vec\Omega\cdot\vec\sigma\right)^{2}|m,+\rangle
-\left(\langle m,+|\vec\Omega\cdot\vec\sigma|m,+\rangle\right)^{2}\nonumber\\ 
& & \quad=1-\frac{\left(2S\right)^{2}}{\left(2S+2\right)^{2}}
=4\frac{2S+1}{\left(2S+2\right)^{2}}\,.
\end{eqnarray}
As a result, the eigenstates $|m,-\rangle$ of the Hamiltonian 
${\cal H}^{\prime}_{(+)}$ are for any system size simultaneously also
eigenstates of ${\cal H}_{(+)}$, whereas the eigenstates $|m,+\rangle$
achieve this property in the thermodynamic limit $S\to\infty$.
In particular the energy $\varepsilon_{-}$ of ${\cal H}_{(+)}$
corresponding to $\varepsilon^{\prime}_{(-)}$ for $n=0$ is just 
$\varepsilon_{-}=-\hbar v/ \ell\sqrt{S}$. 
Moreover, replacing, as an approximation,
the operator $\vec\Omega\cdot\vec\sigma$ with $\pm 1$ one finds for
the energy spectrum of ${\cal H}_{(+)}$
\begin{equation}
\varepsilon_{\pm}\approx\pm\frac{\hbar v}{\ell\sqrt{S}}
\left(n+\frac{1}{2}\mp\frac{1}{2}\right)\,.
\label{ergsp2}
\end{equation}
This result should be seen in analogy to the conventional spherical model
(\ref{halham}). Here the energies increase quadratically with the Landau level 
index $n$, while the spectrum of the underlying planar model is
of course equidistant with energies proportional to $n$.
In the case of graphene, the energies of the planar model are
proportional to $\sqrt{n}$, while in the spherical model ${\cal H}_{(+)}$ they
increase with $\sqrt{n}^{2}=n$.
As seen above, the approximation underlying Eq.~(\ref{ergsp2}) is exact for
both branches of the spectrum at $n=0$ and $S\to\infty$. It is an interesting
speculation whether it becomes also exact in the thermodynamic limit
if $n\neq 0$.

The single-particle states given above were also examined some time ago
by Rezayi in circumstances of quantum Hall skyrmions \cite{Rezayi97}, 
an analogy we shall explore in some detail in section \ref{manybody}.

\subsubsection{Two-body states and interaction matrix elements}

For the conventional spherical model for massful charge carriers 
described in section \ref{haldane}, a particularly simple form
for two-body states of given total angular momentum can be devised
\cite{Haldane83,Fano86}. Using these expressions, matrix elements of
rotationally invariant interactions are conveniently parameterized
in terms of pseudopotentials \cite{Fano86}.

In the present case of massless particles, the pseudospin degree of freedom
adds to the complexity of the two-body problem, and we have not found
a similarly concise expression for states with good quantum numbers
of the total angular momentum. However, what is usually needed in numerical
implementations of interaction Hamiltonians are matrix elements
between tensor products of single-particle states which are related to
pseudopotentials via Clebsch-Gordan coefficients \cite{Fano86}. 
Now, using the expansions
(\ref{psi+}), (\ref{psi-}) it is straight forward to express such
interaction matrix elements of states of massless particles considered here
in terms of those of massful objects given by Eq.~(\ref{halstate}).
Thus, regarding numerical implementations of interaction operators, it is
an easy and straightforward task to adjust an existing code
for the conventional spherical model to the single-particle states of massless
carriers.

\subsubsection{Many-body states}
\label{manybody}

We now discuss many-body states of fully filled energetic sublevels 
(\ref{ergsp}) of lowest Landau level index $n=0$. Let $|\Psi^{\pm}\rangle$
denote the Slater determinant of all $2S+1\pm 1$ single-particle states
given in Eq.~(\ref{psi+}) and  (\ref{psi-}), respectively.
Both many-body states are singlets of the total angular momentum.
For the particle density one naturally finds
\begin{eqnarray}
n^{\pm}(\vec r) & = & \langle\Psi^{\pm}|\sum_{i}
\delta(\vec r-\vec r_{i})|\Psi^{\pm}\rangle\nonumber\\
 & = & \frac{2S+1\pm1}{4\pi\ell^{2}S}\,,
\end{eqnarray}
while the pseudospin density is given by
\begin{eqnarray}
\vec\sigma^{\pm}(\vec r) & = & \langle\Psi^{\pm}|\sum_{i}
\vec\sigma_{i}\delta(\vec r-\vec r_{i})|\Psi^{\pm}\rangle\nonumber\\
 & = & \pm\frac{2S}{4\pi\ell^{2}S}
\left(\begin{array}{c}
\sin\vartheta\cos{\varphi} \\ \sin\vartheta\sin{\varphi} \\ 
\cos\vartheta
\end{array}\right)\,,
\end{eqnarray}
in accordance with Eqs.~(\ref{spinexp+}), (\ref{spinexp-}).
We see that these pseudospin densities form the typical ``hedgehog'' 
structures known from skyrmions \cite{Yoshioka02,Moon95}. However, differently
from skyrmions in conventional quantum Hall monolayers, the physical
electron spin as well as the valley spin are polarized here while
the sublattice spin is forming a topologically nontrivial structure. In fact,
the single particle wave functions (\ref{psi+}), (\ref{psi-}) along with the
many-body states $|\Psi^{\pm}\rangle$ where discussed already in 
Ref.~\cite{Rezayi97} as models for (electron spin) skyrmions. Here we have
identified formally the same skyrmion states (with respect to the 
sublattice spin) as ground states of fully filled Landau levels
of massless Dirac particles. We stress the fact that the Hilbert spaces
spanned by the single-particle states (\ref{psi+}), (\ref{psi-})
for massless Dirac particles are different form the Hilbert space spanned
by the single-particle wave functions (\ref{halstate}) for conventional massful
carriers. In several recent publications however, the Landau levels of graphene
were modeled by the conventional wave functions for massive particles
\cite{Apalkov06,Toke06,Toke07a,Toke07b,Shibata08}.

Finally, in order to evaluate interaction terms within the many-body
states $|\Psi^{\pm}\rangle$ it is useful to consider the
pair distribution function
\begin{equation}
g^{\pm}(|\vec r_{1}-\vec r_{2}|)=\langle\Psi^{\pm}|\sum_{i\neq j}
\delta(\vec r_{1}-\vec r_{i})\delta(\vec r_{2}-\vec r_{j})|\Psi^{\pm}\rangle\,.
\end{equation}
Here we obtain
\begin{eqnarray}
g^{+}(r) & = & \frac{(2S+2)^{2}}{\left(4\pi\ell^{2}S\right)^{2}}
\Biggl[1-\frac{(2S)^{2}}{(2S+2)^{2}}
\left(1-\frac{r^{2}}{2\ell^{2}2S}\right)^{2S-1}\nonumber\\
 & & \qquad\qquad-\frac{4S+2}{(2S+2)^{2}}
\left(1-\frac{r^{2}}{2\ell^{2}2S}\right)^{2S}\Biggr]
\end{eqnarray}
and
\begin{equation}
g^{-}(r)=\frac{(2S)^{2}}{\left(4\pi\ell^{2}S\right)^{2}}
\left[1-\left(1-\frac{r^{2}}{2\ell^{2}2S}\right)^{2S-1}\right]\,.
\end{equation}
From these expressions the ground state energies for arbitrary 
two-body interactions can be evaluated by integration. In particular, for
Coulomb interaction we find
\begin{eqnarray}
E^{+} & = & -\frac{e^{2}}{\epsilon\ell}\frac{1}{\sqrt{S}}
\frac{2^{4S-1}}{\left(
\begin{array}{c}
4S \\ 2S
\end{array}
\right)}
\left(S+\frac{4S+2}{4S+1}\right)\,,
\label{eplus}\\
E^{-} & = & -\frac{e^{2}}{\epsilon\ell}\frac{1}{\sqrt{S}}
\frac{2^{4S-1}}{\left(
\begin{array}{c}
4S \\ 2S
\end{array}
\right)}S\,.
\label{eminus}
\end{eqnarray}
Here $\epsilon$ is the dielectric constant of the host material, and 
we have as usual assumed that the direct (Hartree-) contribution
to the grond state energy is cancelled against a neutralizing background.
The result (\ref{eminus}) agrees with the one given in \cite{Rezayi97}, while
Eq.~(\ref{eplus}) differs in detail form statements made there by 
a contribution which, however, vanishes in the thermodynamic limit.
If the states $|\Psi^{\pm}\rangle$ are interpreted as charged excitations
of a conventional quantum Hall ground state at a filling factor of unity,
their energies should be compared with conventional ground state energy
given by \cite{Schliemann03}
\begin{equation}
E^{0}=-\frac{e^{2}}{\epsilon\ell}\frac{2S+1}{\sqrt{S}}
\frac{2^{4S}}{\left(
\begin{array}{c}
4S+2 \\ 2S+1
\end{array}
\right)}S\,.
\label{enull}
\end{equation}
To provide a meaning full comparison \cite{Moon95}, 
one also has to take into account
that the particle number in the excited states $|\Psi^{\pm}\rangle$ differs
from the conventional ground state by $\pm 1$. Then one finds
for the excitation gap in the thermodynamic limit $S\to\infty$
\begin{equation}
E^{\pm}-\frac{2S+1\pm 1}{2S+1}E^{0}
\to\frac{e^{2}}{\epsilon\ell}\frac{1}{4}
\sqrt{\frac{\pi}{2}}\,,
\end{equation}
in accordance again with Ref.~\cite{Rezayi97},
which is exactly the result predicted by field-theoretical considerations
 \cite{Moon95}. 
Moreover, in Ref.~\cite{Rezayi97} it was also found 
numerically, that the state $|\Psi^{+}\rangle$ has a vanishing variance
for Coulomb interaction operator in the thermodynamic limit, i.e.
this state becomes an eigenstate of the interaction operator in the limit
of an infinite system. This surprising result was also reported for 
other types of long-range interactions\cite{Rezayi97}. 

\section{Conclusions and outlook}
\label{conclusions}

We have introduced  spherical models for the massless Dirac charge carriers
of graphene being subject to a perpendicular magnetic field. 
The Hamiltonians ~(\ref{hamsp}), (\ref{hamsp2})
presented here are analogues of Haldane\rq s spherical 
construction for massful charge carriers. While the first Hamiltonian
(\ref{hamsp}) is arguably closer to the planer model of graphene, 
the latter one (\ref{hamsp2}) can be analyzed easily by elementary 
angular momentum theory. Both Hamiltonians differ by a nontrivial operator.
However, in the subspace of lowest Landau level index $n=0$ the
eigenstates of the single-particle Hamiltonian (\ref{hamsp2}) become 
also eigenstates of (\ref{hamsp}). The latter result holds for the states
of positive energy in the thermodynamic limit, while for the states of negative
energy this statement is true at arbitrary system size.
In particular, the Hilbert spaces
spanned by the single-particle eigenstates of the Hamiltonians
introduced here are different form the Hilbert space spanned
by the single-particle wave functions for conventional massful
carriers. It is a very interesting question for future work, whether
recently reported results of numerical studies of quantum Hall physics 
in graphene using conventional wave functions for massive particles
\cite{Apalkov06,Toke06,Toke07a,Toke07b,Shibata08} are possibly
altered if single-particle states for massless charge carriers 
are used. Indeed, the single particle states constructed here are easily
implemented in existing numerical code for many-body problems in
conventional quantum Hall systems. Finally, the many-body states of
fully filled sublevels in the subspace of lowest Landau level index
are skyrmions with respect to the sublattice spin. 

\acknowledgments{I thank E.~S. Bernardes and J.~C. Egues for useful
discussions and acknowledge the hospitality of the Instituto de Fisica de 
Sao Carlos, University of Sao Paulo, Brazil, where this manuscript reached its
final from. This work was supported by DFG via SFB 689  
``Spin Phenomena in reduced Dimensions''.}


\begin{thebibliography}{99}
\bibitem{note1}
In oder to distinguish this valley index $(\pm)$ from other double signs
occuring later on, we always put it into parentheses.
\end{thebibliography}
\end{document}